\documentclass[fleqn,usenatbib]{mnras}

\usepackage{newtxtext,newtxmath}

\usepackage[T1]{fontenc}

\DeclareRobustCommand{\VAN}[3]{#2}
\let\VANthebibliography\thebibliography
\def\thebibliography{\DeclareRobustCommand{\VAN}[3]{##3}\VANthebibliography}

\usepackage{graphicx}	
\usepackage{amsmath}	
\usepackage{float}	
\usepackage{subcaption}

\newcommand{\Msun}{\mathrm{M}_{\odot}}
\newcommand{\kms}{\text{km} \, \text{s}^{-1}}

\newcommand{\Vlos}{V_\mathrm{los}}
\newcommand{\Vres}{V_\mathrm{res}}

\newcommand{\Vsm}{V_\mathrm{sm}}

\newcommand{\Sint}{S_\mathrm{int}}
\newcommand{\rms}{\sigma_\mathrm{RMS}}

\newcommand{\SN}{S/N}

\newcommand{\AHI}{A_\mathrm{\text{H\,{\sc i}}}}
\newcommand{\Akin}{A_\mathrm{\text{kin}}}
\newcommand{\AHA}{A_{\HA}}

\newcommand{\HI}{\ion{H}{i}}

\newcommand{\HA}{\mathrm{H}{\alpha}}

\newcommand{\FHA}{F_{\mathrm{H}{\alpha}}}

\newcommand{\lgMstarMsun}{\log(M_{\star}/\Msun)}

\newcommand{\lgGF}{\log(M_{\text{\HI}}/M_{\star})}

\newcommand{\Reff}{R_\mathrm{e}}
\newcommand{\cp}{\citep}
\newcommand{\ct}{\citet}

\title[Global asymmetry from $\HA$ to \HI]{SAMI-\HI: the connection between global asymmetry in the ionised and neutral atomic hydrogen gas in galaxies}

\author[A. B. Watts et al.]{
Adam B. Watts,$^{1,2}$\thanks{E-mail: adam.watts@uwa.edu.au}
Luca Cortese,$^{1,2}$
Barbara Catinella,$^{1,2}$,
Chris Power$^{1,2}$, 
Amelia Fraser-McKelvie$^{1,2}$,
\newauthor{Julia J. Bryant$^{2,3,4}$,
Scott M. Croom$^{2,3}$,
Jesse van de Sande$^{2,3}$,
Joss Bland-Hawthorn$^{2,3,4}$,
and Brent Groves$^{1,2}$}
\\
$^{1}$International Centre for Radio Astronomy Research, The University of Western Australia, Crawley, WA, Australia\\
$^{2}$ARC Centre of Excellence for All-Sky Astrophysics in 3 Dimensions (ASTRO 3D), Australia\\
$^{3}$Sydney Institute for Astronomy (SIfA), School of Physics, The University of Sydney, NSW 2006, Australia\\
$^{4}$Australian Astronomical Optics, AAO-USydney, School of Physics, The University of Sydney, NSW 2006, Australia\\
}

\date{Accepted XXX. Received YYY; in original form ZZZ}

\pubyear{2022}

\begin{document}
\label{firstpage}
\pagerange{\pageref{firstpage}--\pageref{lastpage}}
\maketitle

\begin{abstract}
{Observations of the neutral atomic hydrogen (\HI) gas in galaxies are predominantly spatially unresolved, in the form of a global \HI\ spectral line.}
{There has been substantial work on quantifying asymmetry in global \HI\ spectra (`global \HI\ asymmetry'), but due to being spatially unresolved, it remains unknown what physical regions of galaxies the asymmetry traces, and whether the other gas phases are affected.}
{Using optical integral field spectrograph (IFS) observations from the Sydney AAO Multi-object IFS (SAMI) survey {for which global \HI\ spectra are also available (SAMI-\HI)}, we study the connection between asymmetry in galaxies' ionised and neutral gas reservoirs to test {if and how} they can help us better understand the origin of global \HI\ asymmetry}.
{We reconstruct the global $\HA$ spectral line from the IFS observations and find that, while some global $\HA$ asymmetries can arise from disturbed ionised gas kinematics, the majority {of asymmetric cases} are driven by the distribution of $\HA$-emitting gas.}
{When compared to the \HI, we find no evidence for a relationship between the global $\HA$ and \HI\ asymmetry.}
{Further, a visual inspection reveals that cases where galaxies have qualitatively similar $\HA$ and \HI\ spectral profiles can be spurious, with the similarity originating from an irregular 2D $\HA$ flux distribution.}
{Our results highlight that comparisons between global $\HA$ and \HI\ asymmetry are not straightforward, and that many global \HI\ asymmetries trace disturbances that do not significantly impact the central regions of galaxies.}
\end{abstract}




\section{Introduction}
{The physical mechanisms that drive galaxy evolution can leave signatures in the distribution and kinematics of their baryons}. 
{Neutral atomic hydrogen (\HI) gas in galaxies is a useful tracer of these signatures as it is present throughout their optical component and extends to large radii, making it sensitive to both internal and external perturbations \cp[e.g.,][]{vollmer04,angiras06,angiras07,vaneymeren11,vaneymeren11a,cortese21,reynolds21}.}
{The most numerous form of \HI\ observations, however, are not spatially resolved \cp{saintonge22}.}
{Rather, they consist of a {galaxy-integrated} spectral line that is doppler broadened by a galaxy's line-of-sight velocity distribution and weighted by the \HI\ mass distribution.}
{These `global' \HI\ spectra contain combined information about the \HI\ distribution and kinematics of a galaxy, making their shape sensitive to disturbances in both components.}

{Studies of asymmetry in global \HI\ spectra (`global \HI\ asymmetry') indicate that the typical spectrum is not a symmetric profile, with asymmetry rates varying from $35-50$ per cent \cp[depending on the definition,][]{richter94,haynes98,watts20}.}
{Some of this must originate from environmental processes, as galaxies in denser environments host larger global \HI\ asymmetry than more isolated systems \cp[e.g.,][]{scott18,watts20},  and cosmological simulations suggest that this is partly due to the hydrodynamical gas removal and tidal stripping of satellite galaxies \cp{marasco16,yun19,watts20a,manuwal21}.}
{However, this cannot explain the high rate of global \HI\ asymmetries observed in \textit{all} galaxies, regardless of their environment.}
{The origins of these asymmetries remain somewhat unclear, as many studies find significant scatter between global \HI\ asymmetry and galaxy properties \cp{matthews98,haynes98,espada11,reynolds20a,glowacki22}.}
{This makes determining a driving mechanism difficult, and has led to multiple processes being suggested, including gas accretion \cp{matthews98,bournaud05,watts21}, feedback from stars and {active galactic nuclei} \cp{manuwal21}, and tidal interactions and mergers \cp{ramirez-moreta18,bok19,semczuk20,zuo22}.}

{As radio astronomy moves toward Square Kilometer Array \cp[SKA][]{dewdney09}-era surveys, global \HI\ spectra will remain the most numerous form of \HI\ observations.}
{Surveys with SKA precursor telescopes such as  WALLABY\footnote{Widefield ASKAP L-band Legacy Allsky Blind surveY} \cp{koribalski20,westmeier22} are estimated to detect $\sim210\, 000$ global \HI\ spectra below $z<0.1$, compared to $\sim2500$ spatially-resolved sources.}
{Understanding what physical mechanisms are traced by global \HI\ asymmetries is essential to fully exploit the wealth of information contained in these next-generation datasets.}
{Global \HI\ spectra, however, are subject to several limitations.}
{The uncertainty in a global \HI\ asymmetry measurement is a strong function of observational noise \cp{watts20,yu20,deg20}, and the large diversity of physical processes that can drive global \HI\ asymmetry also makes their origin difficult to identify.}
{{These limitations} make it a statistical property that is best used for populations of galaxies \cp{watts20,watts20a,watts21, reynolds20a}, rather than being strongly meaningful for individual objects.}
{Further, {given} their lack of spatial resolution and, considering the typical \HI\ distribution in galaxies \cp{bigiel12,wang14}, it is often assumed that global \HI\ spectra primarily trace galaxy gas reservoirs outside their optical component.} 
{Thus, \textit{it remains unclear whether the disturbances traced by global \HI\ asymmetry are present across the whole galaxy}, or just in the larger radii gas.}

{As spatially resolved \HI\ observations for a large sample of galaxies remain unfeasible outside the very local Universe, one way to progress our understanding of asymmetry in galaxies is to analyse their multi-phase gas reservoirs.}
{The ionised gas, traced through the $\HA$ emission line by integral field spectroscopic (IFS) surveys, is a useful tracer of the dynamics of the baryons in galaxies inside their bright optical components \cp[e.g.,][]{bloom17,bryant19,johnson18}.}
{Moreover, as $\HA$ and \HI\ trace the gas reservoirs of galaxies on different physical scales, this allows the study of the gas dynamics across their whole disc  \cp[e.g.][]{andersen06,andersen09}.}
{There is increasing synergy between \HI\ observations and IFS surveys, such as the \HI\ follow-ups of the MaNGA\footnote{Mapping Nearby Galaxies at Apache point observatory} \cp{bundy15} survey, \HI-MaNGA \cp{masters19,stark21}, and SAMI\footnote{Sydney AAO Multi-object IFS} \cp{croom21} Galaxy Survey, SAMI-\HI\ (Catinella et al., in press, from here C22).}
{It remains unclear, however, what is the connection between \HI\ and $\HA$ asymmetries in galaxies, and whether the two can be meaningfully combined.}

{This paper {presents a pilot study on the combination of $\HA$ asymmetries with global \HI\ spectra using the SAMI-\HI\ survey.}}
{Our goal is to investigate the systematics involved in this analysis and highlight both the complications that arise, as well as the regions of parameter space where the comparison is the most meaningful.}
{This paper is formatted as follows: in \S\ref{sec:data} we describe our sample, and in \S\ref{sec:methods} we describe how we reconstruct the global spectral lines from the IFS observations and our measurement of asymmetry.}
{In \S\ref{sec:results}, we investigate disturbances in the $\HA$ reservoirs and their relationship to the \HI, and in \S\ref{sec:concl} we discuss our results in context with the literature and conclude.}
{The physical quantities used in this work have been calculated assuming a \ct{chabrier03} stellar initial mass function and a $\Lambda$CDM cosmology with $\Omega_\Lambda=0.7$, $\Omega_\mathrm{M}=0.3$, and $h=0.7$.}

\section{Dataset} \label{sec:data} 
The SAMI {galaxy} survey \cp{bryant15,croom12,croom21} is an IFS survey of 3068 galaxies in the stellar mass and redshift ranges $7\leq \lgMstarMsun \leq 12$ and $z<0.1$, respectively. 
It consists of {a} sample of 2180 galaxies that are drawn primarily from the Galaxy And Mass Assembly survey \cp{driver11} as described in \ct{bryant15}, and eight targeted galaxy clusters (888 galaxies) as described in \ct{owers17}. 
The SAMI instrument observed galaxies with {circular} bundles of 61, 1.6 arcsec fibres (hexabundles) that subtend 15 arcsec on the sky \cp{bryant14}. 
The light from the optical fibres was fed into the AAOmega spectrograph \cp{sharp06} and split between blue ($3700-5700\,$\AA) and red ($6300-7400\,$\AA) wavelength ranges with resolving powers $R=1808$ and $R=4304$, respectively.

A datacube was produced for each wavelength range with 0.5-square arcsec spatial pixels (spaxels) that contain the observed spectrum at each position. 
The observations were reduced with the 2{\sc dfdr} software \cp[two-degree field data reduction software,][]{sharp15} and flux-calibrated using observations of standard stars as described in \ct{allen15}.
The {\sc lzifu} \cp[LaZy-IFU,][]{ho16} routine was used to fit the emission lines in each spaxel, which models and subtracts the continuum emission in each spaxel with a penalised pixel-fitting algorithm \cp[pPXF,][]{cappellari04} using the MILES spectral library \cp{vazdekis10} and additional young single stellar population templates from \ct{gonzalezdelgado05}. 
The continuum-subtracted emission lines were fit with {both one- and multi-component} Gaussians, and in this paper, we use the one-component fits to the $\HA$ emission line. 
{At the wavelength of the $\HA$ emission line, the $0.596\,$\AA\ channelisation of the datacubes corresponds to  $\sim 26.8 \, \kms$ channels in line-of-sight velocity.}

{SAMI-\HI\ is an \HI-dedicated {follow-up of a} sub-sample of 296 SAMI galaxies over the stellar mass range $7.4 \leq \lgMstarMsun \leq 11.1$ and redshift $z<0.06$ (C22).}
{The \HI\ observations consist of 143 archival ALFALFA\footnote{Arecibo L-band Fast ALFA} survey \cp{haynes18} detections, and an additional 153 galaxies targeted with the Arecibo radio telescope until the global \HI\ spectral line was detected with an integrated signal-to-noise of at least 5.}
The 3.5 arcmin half-power beam-width of the Arecibo radio telescope is $\sim 14$ times larger than the angular size of the SAMI hexabundles and is sufficiently large to encapsulate the entire \HI\ reservoir of galaxies.
{Stellar mass-estimates for SAMI-\HI\ are derived from GAMA $g$-, and $i$-band photometry using a fitting function that approximates the stellar mass-estimates from full optical spectral energy distribution (SED) modelling \cp{taylor11,bryant15}}.
{Star-formation rates (SFR) are computed from the best-fit model to the UV to FIR SED using {\sc magphys} \cp{dacunha08}, which uses energy balance to derive the obscuration-corrected SFR averaged over the last 100 Myr \cp{davies16a,wright17}.}

\section{Spectrum parameterisation and measurements} \label{sec:methods}
\subsection{Global spectra from \texorpdfstring{$\HA$}{Ha} kinematics}
\begin{figure*}
    \centering
    \includegraphics[width=0.95\textwidth]{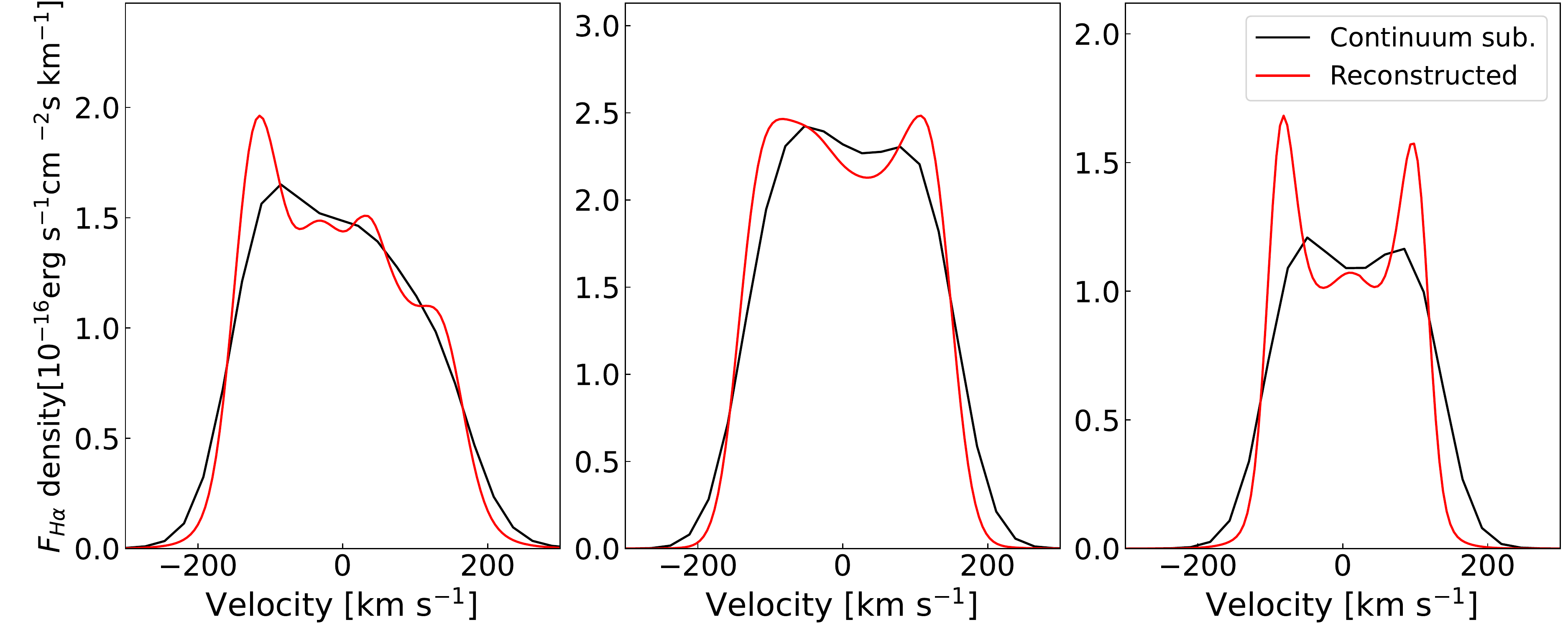}
    \caption[Comparison between continuum-subtracted and reconstructed global $\HA$ spectra]{Comparison between continuum-subtracted (black) and reconstructed global $\HA$ spectra (red) for 3 SAMI-\HI\ galaxies {(catalogue IDs are given in the top-left corner of each panel)}. {Reconstructed spectra better represent the double-horn structure of global emission lines.} \label{fig:native_reconst_spec}}
\end{figure*}

Our sample's global \HI\ observations are not spatially resolved; instead, they represent the \HI\ mass-weighted line-of-sight velocity ($\Vlos$) distribution of the gas reservoir. 
Conversely, the SAMI observations are {resolved}, and while increased spatial resolution in observations is typically advantageous, the two data types are not immediately comparable. 
In particular, disturbance measurements in the 2D $\HA$ flux distribution and kinematics may trace different information than the 1D measure {used} for global \HI\ spectra.
For the same reasons, visual comparison between the global \HI\ spectra and $\HA$ distribution and kinematics is difficult. 
To perform the fairest comparison between $\HA$ and \HI\, we extract global spectra from the SAMI data products that can be analysed in the same way, and thus directly compared to, the global \HI\ spectra. 

{While it is possible to sum the continuum-subtracted datacube across its spatial axes and isolate the $\HA$ emission line (the `continuum-subtracted' global spectrum), we found it was advantageous to reconstruct the {global} emission line using the {spatially-resolved $\HA$ information derived from the {\sc lzifu} fits}.}

{Using the 1-component Gaussian fits to the continuum-subtracted datacube, we distributed the integrated $\HA$ flux in each spaxel $(F_{\HA,ij})$ over a Gaussian centred on its $V_{\mathrm{los,} ij}$ with a standard deviation given by its velocity dispersion ($\sigma_{ij}$). }
{The `reconstructed' global spectrum is computed as the sum of the contributions from all spaxels over a velocity range $v$,}
\begin{equation} \label{eq:HAspec}
S_{\HA}(v) = \sum_i \sum_j \frac{F_{\HA,ij}}{\sigma_{ij} \sqrt{2\pi}} \exp\Big[{-\frac{1}{2}\Big(\frac{v - V_{\mathrm{los,} ij}}{\sigma_{ij}}\Big)^{2}}\Big],
\end{equation}
{in a similar spirit to \ct{andersen06} and \ct{andersen09}.}
To remove the contribution from low-quality spaxels, we required spaxels to have a $\HA$ signal-to-noise ratio $S/N_{\HA}\geq5$, defined as the ratio of the integrated $\HA$ flux to the estimated uncertainty. 

In Fig.~\ref{fig:native_reconst_spec}, we compare the continuum-subtracted and reconstructed global $\HA$ spectra for three SAMI-\HI\ galaxies.
The reconstructed emission lines have the same general shape and width as the continuum-subtracted ones, giving us confidence that they trace the same dynamics. 
However, the double-horn structure in the reconstructed emission lines is clearer, and this is because the reconstruction method reduces two systematic issues with the continuum-subtracted lines.
First, the {width} of the Gaussian fit to the $\HA$ emission line in each spaxel is corrected for instrumental broadening, which smooths out the profile and can suppress and blend peaks into the remainder of the spectrum. 
This effect is also visible in the broader wings of the continuum-subtracted profiles compared to the reconstructed ones. 
{Second, the Gaussian fit to the spectrum in each spaxel provides a more accurate centroid, and can be sampled to finer resolution than the $26.8 \, \kms$ channels of the datacube.}
{Combined with {the corrected} velocity dispersion, this finer sampling of the  $\Vlos$ field of a galaxy makes the reconstructed spectral line {more detailed.}}
To adopt velocity channelisation similar to our \HI\ data, we created our reconstructed spectra with $\Vres=5\, \kms$.

\begin{figure*}
    \centering
    \includegraphics[width=\textwidth]{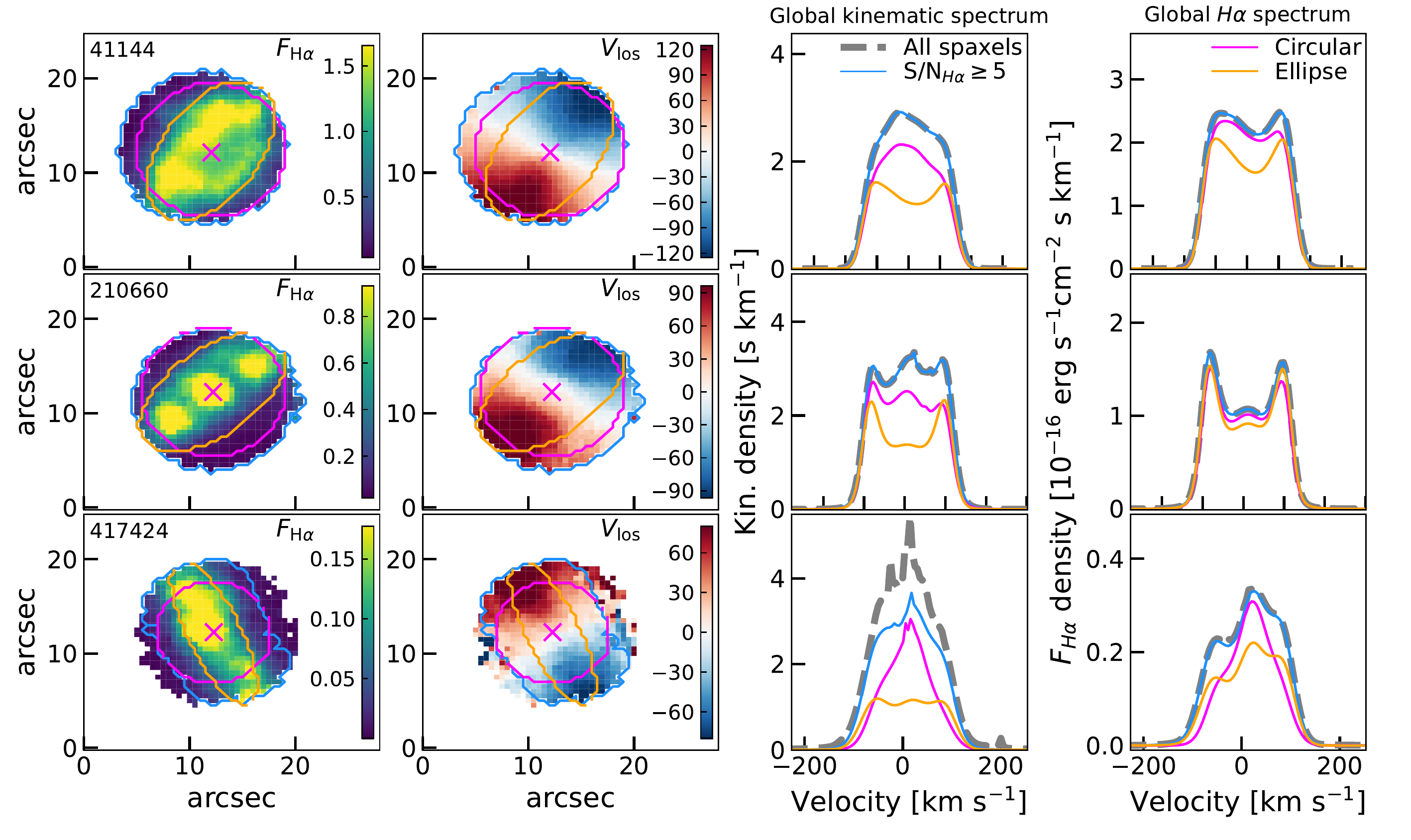}
    \caption[Comparison of global $\HA$ spectra extracted using different apertures]{Comparison of global IFS spectra extracted from different apertures for 3 SAMI-\HI\ galaxies (rows, {catalogue IDs are given in the top left corner of the left panels}). From left to right: $\FHA$ map (units $10^{-16}\,$erg\,s$^{-1}$\,cm$^{-2}$), $\Vlos$ map (units $\kms$), global kinematic spectrum ({not flux weighted}), and global $\HA$ spectrum ({flux weighted}). The maximum value of the $\FHA$ map is set to the $90^\mathrm{th}$ percentile of the $\FHA$ distribution, and in the $\Vlos$ map, the maximum value is the $90^\mathrm{th}$ percentile of the $|\Vlos|$ distribution.
    Colour bars are shown next to each map, and the magenta cross shows our fit $r$-band centre. 
    {Grey, dashed spectra were created using all spaxels, while the coloured spectra use apertures that include a $S/N_{\HA}$ cut of $\geq5$, also shown with contours in the left two panels.}
    {Light blue spectra include all spaxels with $S/N_{\HA}\geq5$, magenta spectra correspond to a circular aperture, and orange spectra to the chosen elliptical aperture.}
    {Using an elliptical aperture is essential for properly sampling the galaxy $\Vlos$ distribution.} \label{fig:aper_compare}}
\end{figure*}

In addition to the spectral line shape, there is a second advantage to the reconstructed spectral lines.
The $\FHA$ distribution in galaxies is typically clumpy, which can create extra structure in the global emission line or even remove the double-horn shape if centrally concentrated. 
This $\FHA$ contribution can be removed by setting $\FHA=1$ for all spaxels in eq. \ref{eq:HAspec}, and a global spectrum that traces only the ionised gas kinematics can be constructed. 
Thus, we created two spectra for each SAMI observation, collectively referred to as `global IFS spectra'.
The first treats all spaxels equally and informs us about the global kinematics of the gas, which we refer to as the `global kinematic spectrum'. 
The second weights each spaxel by its $\FHA$, which we refer to as the `global $\HA$ spectrum' ({as shown in Fig.~\ref{fig:native_reconst_spec}}). 
It contains combined information about the distribution and the kinematics of the ionised gas, analogous to the \HI\ spectrum. 

{The last} systematic effect that we accounted for when creating our global IFS spectra is the different spatial scales of the SAMI and Arecibo observations. 
While the Arecibo beam size is large enough to encapsulate the entire \HI\ reservoir, SAMI observes a {sub-region} of the optical component of the galaxy.
Thus, {except for nearly face-on galaxies,} SAMI's circular field of view samples larger galactocentric radii along a galaxy's minor axis than the major axis. 
As the circular motion of the gas becomes increasingly tangential to the line-of-sight toward the minor axis, using a circular aperture to create a global IFS spectrum can result in excess emission in the centre of the spectrum ($\Vlos=0$), which can destroy the double-horn shape. 

We created our global IFS spectra using only spaxels within an aperture aligned with the galaxy's photometric orientation to account for this. 
We fit a 2-dimensional Gaussian to the $r$-band continuum emission and used the centroid as the aperture's centre. 
The orientation was defined using the photometric position angle and ellipticity measured within one $r$-band effective radius ($\Reff$) from \ct{deugenio21}. 
The aperture was convolved with a 2.5 arcsec Gaussian to model the SAMI point-spread function, and its size was defined as the largest galactocentric radius with complete azimuthal sampling {of spaxels with $S/N_{\HA}\geq5$}.
{This process of aperture definition does not introduce any spurious results or systematic effects into this work, and we have checked that our results remain qualitatively unchanged if the IFS spectra are created using the entire SAMI {field of view}.}

In Fig.~\ref{fig:aper_compare}, we show how the shape of our global IFS spectra change when using all spaxels {(grey, dashed spectra)}, all spaxels with $S/N_{\HA}\geq5$ (blue apertures and spectra), the subset of the $S/N_{\HA}\geq5$ spaxels within the largest circular aperture (magenta apertures and spectra), and our final elliptical aperture (orange apertures and spectra). 
{The global spectra created using all spaxels agree well with the spectra created using only those with $S/N_{\HA}\geq5$ for the top rows, whereas this is not the case for the galaxy in the bottom row.}
{This object (SAMI ID 417424) has large ellipticity ($e=0.711$), meaning that there are spaxels with low signal off of the galaxy plane on the minor axis that have uncertain kinematics.}
{Not only does this highlight where the $\SN_{\HA}$ cut is important, but the agreement between the grey and blue lines in the other five spectra also explains why we observe no qualitative difference in our results when the IFS spectra are created using the entire SAMI field of view.}

{Considering the aperture spectra,} using either {just the $S/N_{\HA}\geq5$ aperture} or a circular one results in excess flux density at $\Vlos=0$,  {which is} most evident in the global kinematic spectra where each spaxel is weighted evenly. 
In contrast, in the global $\HA$ spectrum, the $\FHA$ weighting typically up-weights the major axis and down-weights the minor axis, recovering more of the double horn shape.
However, the double-horn shape is recovered in both the IFS spectra when using an elliptical aperture.

\subsection{Spectrum measurements} \label{subsec:measure}
{We parameterised our global \HI\ spectra with the busy function \cp{westmeier14}.}
{As our spectral measurements are computed from the data rather than the fits, and we are not interested in parameterising the more complex shape of the central trough, we only needed to determine the locations of the profile edges.}
{Thus, the fits were optimised to describe the edges of the spectrum only \cp[e.g.,][]{watts21}.}
{We visually inspected each fit to ensure the edges were accurately parameterised, and we boxcar-smoothed spectra when necessary.}
{Upper- ($V_U$) and lower- ($V_L$) velocity limits for the global \HI\ spectra were defined to be where the edges of the profile equal twice the RMS measurement noise ($2\rms$), measured in the signal-free part of the spectrum \cp[e.g.][]{watts21}. }
{The integrated flux $(\Sint)$ was measured between these limits, and the integrated signal-to-noise ratio ($\SN$) calculated using}
\begin{equation}
\SN = \frac{\Sint}{\omega\, \rms} \sqrt{\frac{1}{2} \frac{\omega}{\Vsm}},
\end{equation}
{where $\omega= V_U - V_L$ is the velocity width and $\Vsm$ is the final velocity resolution after boxcar smoothing \cp[e.g.][]{saintonge07,watts20}.}

{The absence of per-channel measurement noise} in the global IFS spectra means that we cannot define velocity limits in the same way as the \HI.
Instead, we determined the location of the peak(s) in each global IFS spectrum and defined the velocity limits to be where the edges of the profile equal 20 per cent of the peak flux(es).
We have already included a $\SN$ cut in our IFS spectra through the selection of spaxels with $\HA$ signal-to-noise $\geq5$, and \ct{watts21} showed that velocity limits defined at 20 per cent of the peak flux(es) and $2\rms$ are comparable.

\begin{figure*}
    \centering
    \includegraphics[width=\textwidth]{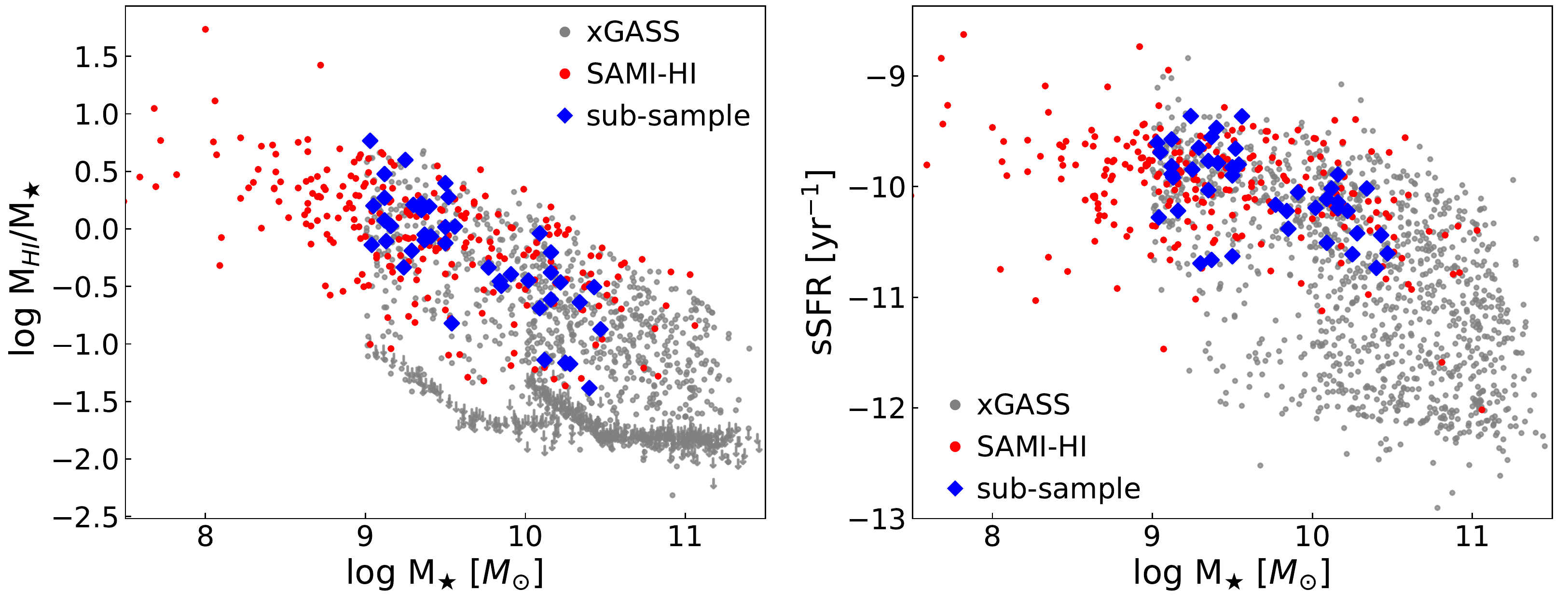}
    \caption[Comparison of the SAMI-\HI\ parent and sub-samples]{Comparison of the SAMI-\HI\ parent and sub-samples. The left panel shows \HI\ mass fraction as a function of stellar mass, and the right panel shows sSFR as a function of stellar mass. In both panels, the SAMI-\HI\ parent sample galaxies are shown as as red points and the final sub-sample as blue diamonds. xGASS galaxies are shown as grey points in the background, {with \HI\ non-detections shown {as} downward arrows}. {The final sub-sample is more restricted in $\lgGF$ and sSFR range than the parent sample.} \label{fig:SHI_parent_final_compare}}
\end{figure*}
\begin{figure*}
    \centering
    \includegraphics[width=\textwidth]{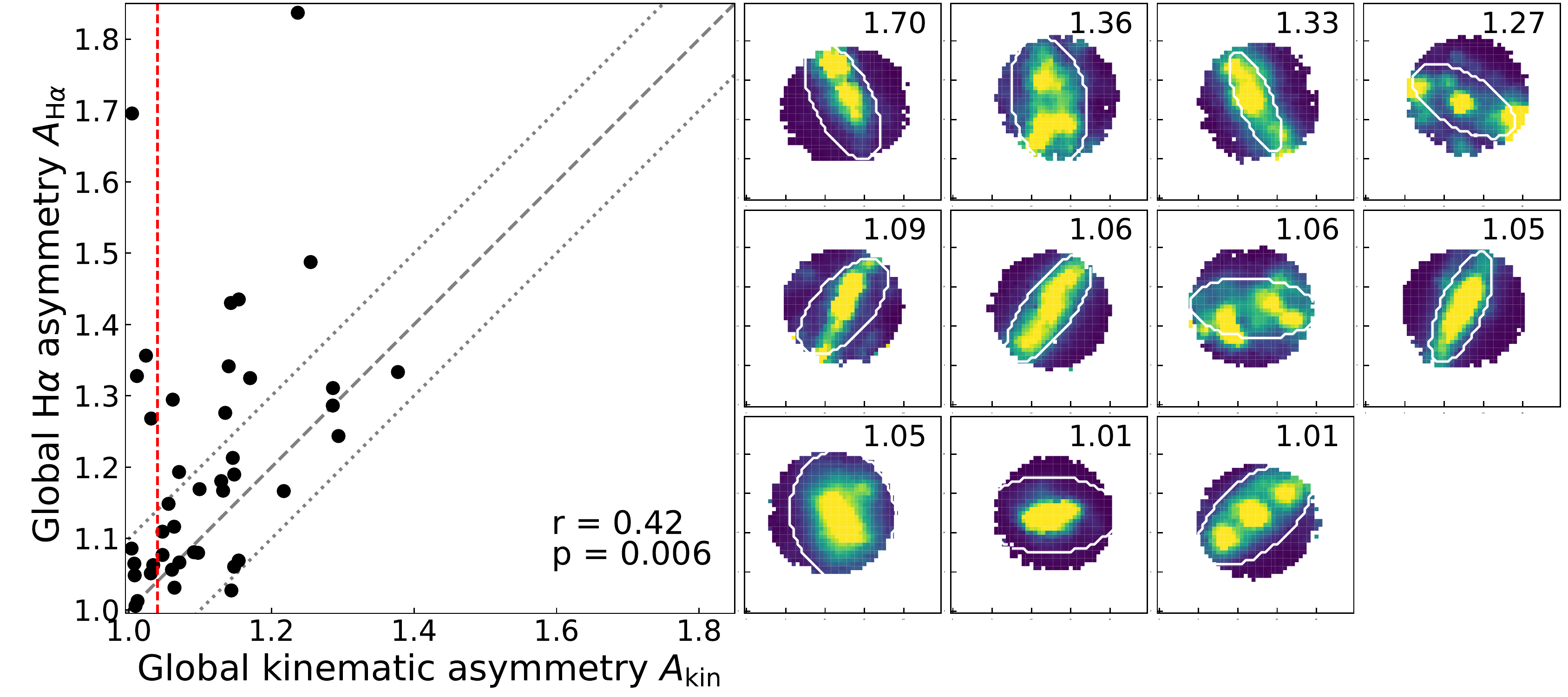}
    \caption[Comparison of global kinematic to $\HA$ asymmetry and the $\HA$ flux distribution]{Left: Comparison between global kinematic and $\HA$ asymmetry. The one-to-one relation is shown with a dashed, grey line and a scatter of $\pm0.1$ by dotted, grey lines. {The Spearman's rank correlation coefficient (r$=0.42$) and significance (p$=0.006$) are shown in the bottom right corner.} The vertical red dashed line corresponds to $\Akin =1.04$, used to select the galaxies on the right. Right: $\HA$ flux distributions for galaxies to the left of the $\Akin=1.04$ line. The galaxies are ordered by their $\AHA$ value, given in the top right of each panel, and the apertures used to extract the global IFS spectra are shown in white. {Elevated $\AHA$ is primarily driven by disturbances in the $\HA$ distribution, rather than disturbances in the kinematics.} \label{fig:AhaAkin_compare}}
\end{figure*}

{Using the velocity limits for each spectrum, we measured the integrated flux ratio asymmetry parameter $A$ \cp[e.g.][]{haynes98,espada11} defined as}
\begin{equation} \label{eq:Afr}
    A = 
        \begin{cases}
            R & R \geq 1\\
            1/R & R < 1
        \end{cases},
\end{equation}
where 
\begin{equation} \label{eq:A}
    R = \frac{
        \int_{V_{M}}^{V_{U}} S_{\mathrm v} {\mathrm dv}
        }{
        \int_{V_{L}}^{V_{M}} S_{\mathrm v} {\mathrm dv}
        },
\end{equation}
and $V_M = 0.5(V_L+V_U)$ is the middle velocity of the spectrum and $S_\mathrm{v}$ is the \HI\ or $\HA$ flux density. 
{The inversion in eq. \ref{eq:Afr} removes the dependence of the value on the approaching/receding half of the spectrum, such that $A=1$ indicates a symmetric spectrum, while $A>1$ indicates a deviation from symmetry in the distribution and/or kinematics of the gas.}
Thus, we have a global \HI\ asymmetry, global kinematic asymmetry, and global $\HA$ asymmetry measurement for each galaxy, which we denote as $\AHI$, $\Akin$ and $\AHA$, respectively. 
{$\AHA$ and $\AHI$ are both flux weighted quantities, making them sensitive to asymmetry in both the distribution and the kinematics of the gas.}
{Conversely, $\Akin$ is not flux weighted, making it sensitive to only kinematic asymmetries in the gas.}

{We note here that we expect the uncertainty due to measurement noise to be largest for $\AHI$, as we have used SAMI spaxels with $\HA$ $\SN\geq 5$ to create our global IFS spectra and measure $\Akin$ and $\AHA$.}
{The uncertainty in $\AHI$ is a sensitive function of $\SN$  \cp[see][for more details]{watts20,deg20,watts21,bilimogga22}, and for our final sample (defined below) the typical absolute uncertainty in an $\AHI$ measurement is $<0.2$ \cp[e.g.][]{yu21}, although it could be as large as $0.35$ in the lowest $\SN$ spectra.}
{We have checked that our results are not sensitive to the $\SN$ distribution of our \HI\ spectra.}

{Asymmetry in global spectra can also be sensitive to the central velocity \cp[e.g.,][]{deg20}, and our method measures this value independently for each global spectrum.}
{We computed the difference in central velocities between the IFS spectra and the global \HI\ spectrum using galaxies that are not \HI-confused and have \HI\ $\SN\geq7$ (see our final sample selection in the next subsection).}
{In both cases we found  a Gaussian distribution with standard deviation of $10\,\kms$, or two channels in our global IFS spectra, implying that the global spectra trace the same overall potential and we can ignore small differences in the central velocity.}

{Last, we note that we have not included internal extinction corrections to the observed $\FHA$ when creating our global $\HA$ spectra.}
{Extinction correction is not always feasible across galaxies as it requires using the H$\beta$ line, which is typically lower $\SN$.}
{Regardless, we explored their effect on our global $\HA$ spectra, and describe our tests in Appendix \ref{app:extcorr}.}
{In summary, the typical scatter between uncorrected and corrected $\AHA$ is $\sigma=0.092$, with no preference toward larger or smaller asymmetry.}

\subsection{Final sample}
To ensure that artefacts do not impact our data, and so our global spectra are comparable, we removed galaxies from our sample with several quality cuts.
We selected 216 galaxies with $\lgMstarMsun \geq 9$ to ensure that our sample remains primarily {rotation-dominated \cp[$\HA$ $V/\sigma=2-3$,][C22]{bloom17a,aquino-ortiz20}.}
{The sample was assessed for \HI\ confusion to remove systems that could have non-negligible contributions from sources outside the target galaxy, as described in C22, which removed an additional 20 galaxies where confusion was certain.}
We adopted an \HI\ $\SN$ cut of $\SN\geq7$, which removed 62 galaxies with low-quality \HI\ spectra that could not be meaningfully compared to the IFS spectra, {and to minimise the impact of measurement noise on our measured $\AHI$}.
From the remaining 134 galaxies, we removed {four} that had SAMI {observations with missing data that caused part of the galaxy to be masked, such that the global IFS spectra did not {fully sample the observed region.}}

Last, we made a cut {based on the radial extent} (i.e., in the number of $r$-band $\Reff$) of our global IFS spectra apertures, as to make a fair comparison between our IFS and \HI\ spectra, they need to trace similar dynamics. 
{Using the whole SAMI-\HI\ sample, C22 showed that the ratio between \HI\ and $\HA$ (measured at 1.3\,$\Reff$) velocity widths decreases from roughly unity at $\lgMstarMsun=10$ to $\sim1.3$ at $\lgMstarMsun=9$.}
{This decrease is due to the mass dependence of rotation curve shapes, such that for low-mass galaxies, the SAMI field of view may not reach the flat part of the rotation curve.}
{To minimise this bias while keeping a reasonable sample size, we focus our analysis on the 63 galaxies with IFS apertures that covered at least $1.5\, \Reff$.}
{This aperture cut means, that, in the worst case scenarios we are missing the last 30 per cent of the rising part of the rotation curve, or $\sim30\,\kms$ (6 channels of our global IFS spectra) in the case of the lowest mass galaxies in our sample.}
{For reference, a minimum aperture of 1\,$\Reff$ includes 104 galaxies (26 removed), but there will be numerous galaxies with a larger discrepancy between their \HI\ and $\HA$ velocities, while an aperture size of 2\,$\Reff$ would result in only 34 galaxies remaining (94 removed).}

{In addition to this aperture cut,} we also excluded a final 21 galaxies where both IFS spectra  were Gaussian-shaped and could not be interpreted in terms of a differentially-rotating disc.
{These objects were typically face-on galaxies, or galaxies where the photometric position angle was significantly offset from the kinematic position angle.}
This left 42 galaxies in our final sub-sample.

{This reduction in sample size highlights the narrow overlap in parameter space between current multi-object IFS surveys and global \HI\ surveys, even in star-forming galaxies.}
{We acknowledge that the small sample size {and restriction to \HI-normal star-forming galaxies (see Fig.~\ref{fig:SHI_parent_final_compare} and the following paragraph) means that it is not a complete, representative sample}, and thus we do not aim to make strong statistical statements about asymmetry in galaxies.}
{Instead, we investigate how disturbances manifest in these different tracers and how they might be meaningfully combined.}
{On this note, we checked for the presence of AGN in our sample, which could contribute bright emission in the centre of galaxies and complicate the interpretation of our global $\HA$ spectra.}
{Using a N{\sc ii}-$\HA$ Baldwin, Phillips, and Terlevich diagram \cp{baldwin81} with spaxels within $0.5\, \Reff$ and $S/N>5$ in all lines, 16 SAMI-\HI\ galaxies had spaxels above the \ct{kewley01} demarcation line.}
{All these galaxies are already removed from our final sub-sample by the other quality cuts listed above, so our results are free from AGN contamination.}

The properties of the final sub-sample are compared to the SAMI-\HI\ parent sample in Fig.~\ref{fig:SHI_parent_final_compare}.
{For context, in Fig.~\ref{fig:SHI_parent_final_compare} we also compare these SAMI-\HI\ samples to xGASS\footnote{extended GALEX Arecibo SDSS Survey} \cp{catinella18}, a deep \HI\ survey that is representative of the \HI\ properties of galaxies in the local Universe.}
{The SAMI-\HI\ sub-sample spans a similar range of \HI\ fractions as the parent sample at $\lgMstarMsun>10$ but has fewer galaxies with $\lgGF<-0.5$ below $\lgMstarMsun<10$, due to the $\SN$ cut limiting the smaller \HI\ masses.}
{It is also clearly restricted to the star-forming main-sequence compared to the SAMI-\HI\ parent sample and contains no galaxies with $\lgMstarMsun\geq10.5$ due to the $1.5\, \Reff$ minimum aperture requirement.}

\begin{figure*}
    \centering
    \includegraphics[width=\textwidth]{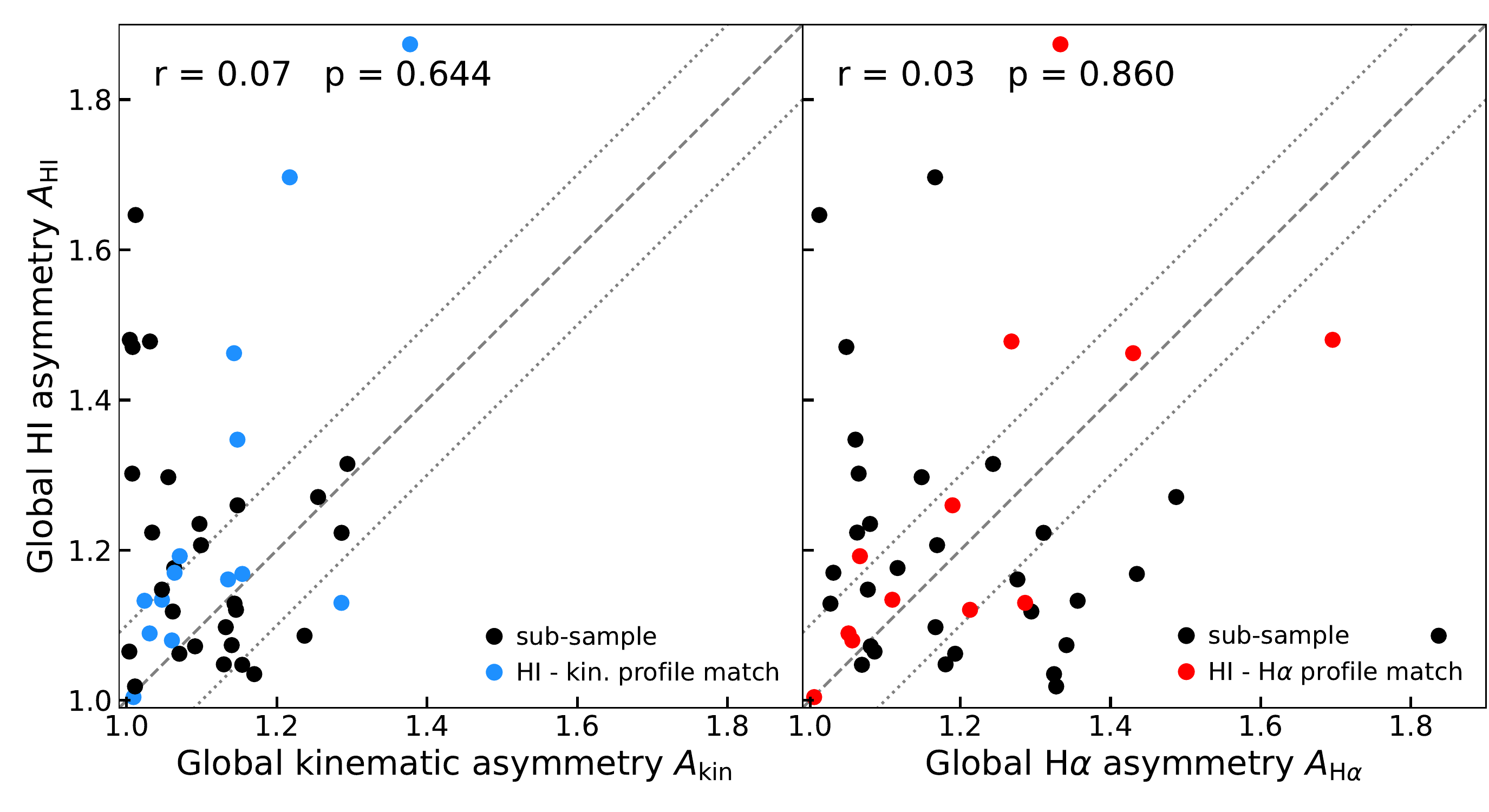}
    \caption[Comparison of global kinematic and $\HA$ asymmetry to global \HI\ asymmetry]{Comparison between global IFS and global \HI\ asymmetry measurements. The left panel compares $\Akin$ and $\AHI$, while the right one compares the $\AHA$ to $\AHI$.  Visually classified matches between the global kinematic and \HI\ spectra are shown as blue points in the left panel, and matches between the global $\HA$ and \HI\ spectra are shown as red points in the right panel. The one-to-one relationship is shown in both panels with a dashed, grey line, and a scatter of $\pm0.1$ with dotted grey lines are {shown for easier comparison with Fig.~\ref{fig:AhaAkin_compare}}. {In the top-left corner of both panels, we give the Spearman's rank correlation coefficient and significance for all points in the panel in black.
    }\label{fig:AifsAhi_compare}}
\end{figure*}
\section{Results}  \label{sec:results}
\subsection{Asymmetry in global IFS spectra}
{In the left half of Fig.~\ref{fig:AhaAkin_compare}, we compare the asymmetry in the global kinematic and $\HA$ spectra, $\Akin$ and $\AHA$.}
The one-to-one relation is shown with a grey dashed line, and the grey, dotted lines demonstrate a scatter of $\pm0.1$ {to guide the eye, but this value is  similar to the scatter observed from our extinction correction tests in Appendix \ref{app:extcorr}.}
{The two asymmetry measurements are moderately correlated, with a Spearman's rank correlation coefficient $r=0.42$ ($p=0.006)$.}
{We note that {the} dynamic range in $\AHA$ is larger than the range in $\Akin$,} and while some galaxies lay inside the $\pm0.1$ scatter, there is a clear extension of points with elevated $\AHA$ {across the whole range of $\Akin$.}

This {preferential {increase} of $\AHA$ at fixed $\Akin$} suggests that the magnitude of disturbance in the kinematics sets a lower limit to the measured asymmetry value, {and the $\FHA$ distribution only increases} the measured asymmetry.
{In other words, the primary cause of asymmetry in the global $\HA$ spectra is asymmetry in the $\FHA$ distribution, rather than kinematic asymmetries in the velocity field, at least from a statistical perspective.}
{This interpretation is supported by the right half of Fig.~\ref{fig:AhaAkin_compare}, where we show the $\FHA$ distribution of galaxies with  $\Akin<1.04$ ordered by decreasing $\AHA$.}
{Galaxies with larger $\AHA$ typically have more irregular $\FHA$ distributions within their elliptical aperture (i.e., within the white ellipse) compared to those with small $\AHA$.}
{This {connection between an irregular $\FHA$ distribution and elevated global $\HA$ asymmetry} agrees with previous results presented by \ct{andersen09}.}

\subsection{Comparison with global \texorpdfstring{H\,{\sc i}}{HI} asymmetry}

In Fig.~\ref{fig:AifsAhi_compare}, we compare $\Akin$ and $\AHA$ to $\AHI$. 
In the left panel, which compares $\Akin$ and $\AHI$, we see a similar distribution to Fig.  \ref{fig:AhaAkin_compare}. 
{There is a trend for $\Akin$ to be generally smaller than $\AHI$, but there is visibly larger scatter than between $\Akin$ and $\AHA$, and the Spearman's rank correlation coefficient $r=0.07$ with $p=0.64$ implies no correlation.}
In the right panel of Fig.~\ref{fig:AifsAhi_compare}, we compare $\AHA$ and $\AHI$, which is observationally the most like-for-like comparison as both are {measured from} flux-weighted spectra.
{There are more points below the lower dotted scatter line compared to the left panel, and the correlation coefficient $r=0.03$ with $p=0.86$ indicates no trend.}
{Thus,} galaxies can have asymmetric global \HI\ spectra and undisturbed global $\HA$ spectra, and vice-versa \cp[e.g.,][]{andersen09}.

\begin{figure*}
    \centering
    \includegraphics[width=0.95\textwidth]{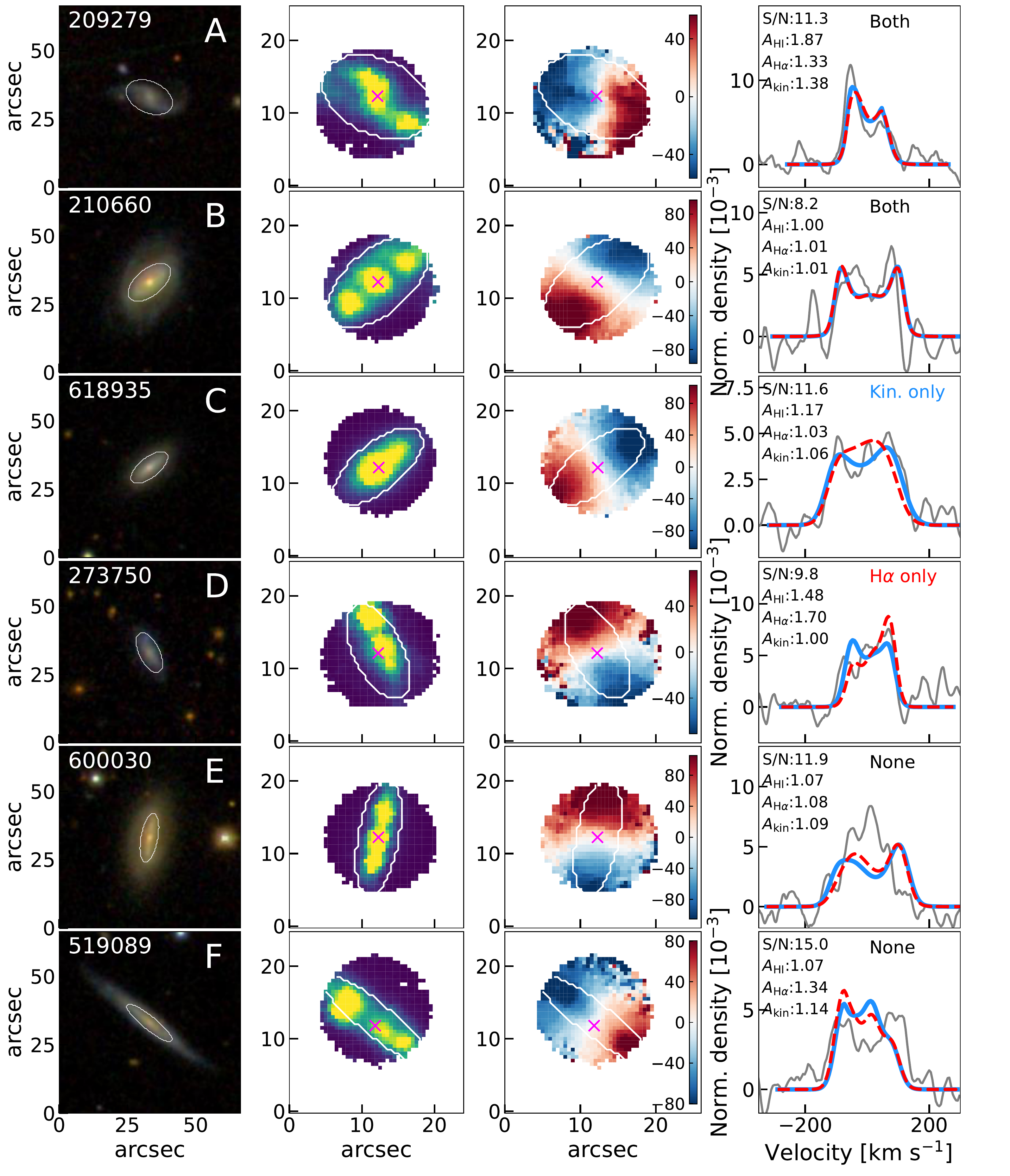}
    \caption{Example SAMI-\HI\ galaxies ordered by matches between global \HI\ and IFS spectra, {with SAMI catalogue IDs given in the top left corner of each optical image}. Left to right, 1x1 arcmin SDSS cutout, integrated $\FHA$ map, $\Vlos$ field with colour bar in $\kms$, and overlaid global \HI\ (grey), kinematic (blue), and $\HA$ (red, dashed) spectra.  The magenta cross shows our fit $r$-band continuum centre, and the white contours indicate the {aperture used for spectrum extraction}. {For scale, the FWHM of the $\sim3.5$ arcmin Arecibo beam would be drawn $\sim1.25$ arcmin beyond the edge of each SDSS cutout.} 
    The \HI\ $\SN$ and the $\AHI$, $\Akin$ and $\AHA$ values are shown in the top left of each spectrum panel, and the text in the top right indicates which IFS spectra were identified to match the \HI. \label{fig:examples}}
\end{figure*}

{Previous studies have noted that the $A$ flux ratio asymmetry measurement can be very sensitive to measurement noise \cp[e.g.][]{watts20,yu20}, particularly at low $\SN$.}
{We have checked that differences in profile $\SN$ cannot explain these poor correlations in our sample.}
{The galaxies with the largest $\AHI$ do not have preferentially low $\SN$, and the data remained uncorrelated when using the subsets with $\SN$ above and below the sample median.}
{Does this imply that there is never a match between the asymmetry in the two gas phases?}
{The presence of galaxies along the one-to-one relation suggests that this is not the case, but we must also take into account the different dynamic ranges of each asymmetry measurement.}
{To make sure {that} we identified real cases where the asymmetry in the two gas phases matches, we visually inspected the global spectra of each galaxy.}
{The spectra were overlaid after being normalised by their integrated flux, centred on their $V_M$, and the \HI\ spectrum boxcar-smoothed to a resolution of $\Vsm=29\, \kms$ {to better match the typical velocity dispersion of the $\HA$ line} ({see Fig.~\ref{fig:examples} for examples}).}
{Each galaxy was assessed by authors ABW and LC independently for a qualitative match between each of the global IFS spectra and the \HI\ spectrum based on the profile width, the relative height of the peaks, and the shape of the central trough.}
{Cases of disagreement were discussed, and we only consider galaxies to be real matches if both assessors agreed.}

{These matching cases are identified in Fig.~\ref{fig:AifsAhi_compare}  using blue points in the left panel (i.e., galaxies with matching global \HI\ and global kinematic spectra), and red points in the right panel (i.e., galaxies with matching global \HI\ and global $\HA$ spectra).}
{Interestingly, these galaxies are not preferentially located in any part of either parameter space, and they span the entire range of each asymmetry measurement.}
{By construction, they have smaller scatter than the whole population (i.e., $r=0.69$ for $\Akin-\AHI$ and $r=0.81$ for $\AHA-\AHI$).}
{However, not every galaxy along the one-to-one line was identified to have a matching global IFS or \HI\ spectrum, even when accounting for the super-linear correlation of the blue points in the  $\Akin-\AHI$ parameter space.}
{This result shows that, while galaxies can have matching asymmetry between the two gas phases, this cannot easily be judged from the quantitative asymmetry alone.} 

{We {further} investigate {{different cases of} spectral profile matches} in Fig.~\ref{fig:examples}, where we show Sloan Digital Sky Survey \cp{york00,abazajian09} images, $\FHA$ and $\Vlos$ maps, and overlaid global \HI\ (grey), kinematic (blue) and $\HA$ (red, dashed) spectra for galaxies with different matching classifications.}
{The first two rows (galaxies A and B) show cases where both global IFS spectra were identified to match the global \HI\ spectrum.}
{Galaxy A appears to host a galaxy-wide disturbance; the optical image shows that it has strong spiral arms, {giving it an elongated appearance}, and all three global spectra match closely.}
{The global spectra for galaxy B are all symmetric, double-horn profiles.}
{While the bar in this galaxy dominates the photometric position angle, the $\FHA$ distribution is axisymmetric, and the kinematics are undisturbed.}
{In these galaxies, it seems that the ionised gas traces the same {kinematics and spatial distribution} as the \HI.}

{In the third and fourth rows (galaxies C and D) of  Fig.~\ref{fig:examples}, we show galaxies where only one of the IFS spectra match the global \HI\ spectrum.}
{Galaxy C has a global kinematic spectrum that matches the \HI, indicating that the kinematics of the two phases might be similar, but the centrally-concentrated $\FHA$ removes the double-horn structure in the global $\HA$ spectrum.}
Conversely, galaxy D has a symmetric global kinematic spectrum that does not match the \HI, but the off-centre $\FHA$ distribution {causes a highly asymmetric global $\HA$ spectrum that appears to better match the \HI.}
{Indeed, the relative spatial distributions of both the $\HA$ and \HI\ could be lopsided in this galaxy while their kinematics are symmetric.}
{Alternatively, this could just be an offset or clumpy $\FHA$ distribution in the galaxy with no connection to the \HI\ distribution.}
{Our inability to decompose the \HI\ spectrum into kinematic and flux-weighted components complicates this interpretation, and although we have categorised this galaxy as an $\HA-$\HI\ match based on its shape,  the low $\SN$ of the \HI\ spectrum means that there is a chance that it could also be consistent with the global kinematic spectrum.}
{Thus, we emphasise that comparisons between global IFS and \HI\ spectra must be interpreted carefully, and that matching profile asymmetries or shapes do not imply that the distribution or kinematics of the $\HA$ and \HI\ are similar, nor do mismatched asymmetries imply that they disagree.}
{In the last two rows (galaxies E and F), there is no match between the global IFS spectra and the \HI.}

{This diversity of $\FHA$ distributions} helps explaining the absence of a correlation between $\AHI$, $\Akin$ and $\AHA$ in Fig.~\ref{fig:AifsAhi_compare}. 
{In addition to the statistical nature of asymmetry measurements that make them less meaningful for individual galaxies, further scatter is introduced by the $\FHA$ distribution, which can cause spurious (mis-)alignments between the shapes of the global \HI\ and IFS spectra.}
{{These results} suggest that, in our sample, the disturbances traced by the global \HI\ spectra must be located outside our IFS apertures or be too weak to cause measurable changes in their inner regions.}
{In the same vein, the processes driving asymmetry in the global $\HA$ spectra do not drive significant asymmetry in the \HI.}
{This interpretation is supported by comparing the physical size of our IFS apertures to the size of the \HI\ disc ($R_{\text{\HI}}$), using the \HI\ size-mass relation from \ct{wang16}}.
{The IFS apertures cover between $0.1-0.6\, R_{\text{\HI}}$, and neither of the visually-identified matching populations cover preferentially larger $R_{\text{\HI}}$.}

\section{Discussion and conclusions}  \label{sec:concl}
In this paper, we have combined IFS observations of $\HA$ emission and global \HI\ spectra of galaxies to study {the asymmetry of their multi-phase gas reservoirs.}
By reconstructing global emission lines from the IFS observations (global IFS spectra) and defining global asymmetry as the ratio between the approaching and receding halves of the spectrum, we have investigated disturbances in the distribution and kinematics of the $\HA$ emission and connected this to the \HI. 

{We found that the distribution of $\HA$ emission almost always determines the asymmetry of a global $\HA$ spectral line, while asymmetry in the $\Vlos$ field typically sets the lower limit.}
Galaxies can have clumpy, asymmetric $\HA$ flux distributions and global $\HA$ spectra, but regular kinematics, similar to the disturbed appearance but regular kinematics exhibited by \HI\ in low mass galaxies \cp[e.g.,][]{hunter12}.
There is no clear correlation between global asymmetry in the $\HA$ distribution or kinematics and global asymmetry in the \HI.  
{This lack of correlation implies that the physical mechanism(s) responsible for driving the global asymmetry in each gas phase are different, and do not affect the other phase. }
{However, it does not mean that the global \HI\ and IFS spectra of a galaxy cannot have the same shape, only that these cases are hard to identify and must also be interpreted carefully.}

This paper has been presented primarily as a description of methodology, {and an investigation into the systematics of how IFS observations can be used} to interpret asymmetry in global \HI\ spectra. 
Our results, however, are also consistent with previous literature.
{{Similar} reconstruction of global $\HA$ spectra and their comparison to global \HI\ spectra was presented in \ct{andersen06} and \ct{andersen09}, and these studies also concluded that the $\FHA$ distribution dominates the asymmetry in global $\HA$ spectra.}
{Further, \ct{andersen09} suggest that the $\HA$ and \HI\ share similar kinematics, and that the poor correlation between the asymmetry of each global spectrum is due to the individual, and differently distributed, flux distribution of each gas phase.}
{While our data confirm that the $\HA$ flux distribution typically dominates the asymmetry of the global $\HA$ spectra, we also find poor correlation between asymmetry in the $\HA$ velocity field and the \HI\ asymmetry.}
{We cannot rule out that this originates from asymmetric \HI\ flux distributions, but the persistence of the poor correlation at small global \HI\ asymmetry suggests that, at least in our IFS apertures, the kinematics of the $\HA$ gas may be different to that of the \HI\ in some cases.}

{These results imply that most global \HI\ asymmetries trace disturbances that do not significantly impact the inner regions of galaxies,  and this can explain the diversity in the optical properties of galaxies with asymmetric global \HI\ spectra previously noted by several studies \cp[e.g.,][]{espada11,watts20,watts21}.}
{Indeed}, the strongest correlations between \HI\ asymmetry and galaxy properties found so far have been with \HI\ content and galaxy environment \cp[e.g.,][]{bok19,watts20,reynolds20,glowacki22}, as these properties are sensitive to processes outside the optical disc of galaxies.
{This absence of a connection between asymmetry in the two gas phases also suggests that mechanisms such as feedback from supernovae or active galactic nuclei (affecting primarily the central $\HA$) or weak gravitational interactions (affecting primarily the distant \HI) are not likely able to cause a coherent global asymmetry in both the $\HA$ and \HI\ gas.}
{Instead, it likely requires repeated or extended perturbation to the whole galaxy that has occurred within the last $\sim2$\,Gyr \cp{bloom18,feng20,ghosh22,lokas22}.}

{It is important to mention the limitations in our analysis, the foremost of which is} the low sample numbers {and the requirement of high $\SN$ spectra to make a meaningful comparison between global \HI\ and IFS spectra.}
{One way to expand this analysis further would be to perform targeted {IFS} observations of galaxies with asymmetric, high signal-to-noise ($\SN\gtrsim20$) global \HI\ spectra.}
{While this would be interesting for individual cases, however, the number of galaxies with high $\SN$ \emph{and} significant asymmetry is not likely to build a statistical sample.}
{Number statistics aside, the reader should keep in mind the differences in the spatial scales traced by the IFU and \HI\ observations, although we found no preference for galaxies that cover a larger fraction of the \HI\ reservoir to be more likely to have matching global IFU and \HI\ profile shapes.}
{This absence of matching profile shapes could also be partly due to the fact that the global \HI\ spectra cannot be decomposed into kinematic and flux-weighted components as was done for $\HA$ observations}.
{This decomposition of the \HI\ would allow for direct comparison between the kinematics of each gas phase, and quantification of how their individual flux-weightings change the measured asymmetry.}
{However, using spatially resolved data will be the best way to understand asymmetries in the multi-phase gas reservoirs of galaxies}.

{{Thus,} next-generation surveys for \HI\ emission, such as WALLABY \cp{koribalski20}, are a promising outlook for studying asymmetry between different gas phases.}
{WALLABY will produce resolved \HI\ observations for {thousands} of galaxies, and while \HI\ {is the best tracer of} the gas at large galactocentric radii, IFS follow-up, or overlap with next-generation IFS surveys such as Hector \cp{bryant20}, will allow {asymmetry in both the gas distribution and kinematics} to be measured in a spatially resolved way {and for a statistical sample of galaxies}.}
{These analyses will provide the best constraints on the origin of, and connection between, disturbances in the multi-phase gas reservoirs of galaxies.}

\section*{Acknowledgements}
We thank the referee for their constructive comments that improved this manuscript. 
ABW acknowledges that part of this work was supported by an Australian Government Research Training Program (RTP) Scholarship.
LC and ABW acknowledge support from the Australian Research Council Discovery Project  funding scheme (DP210100337). 
LC is the recipient of an Australian Research Council Future Fellowship (FT180100066) funded by the Australian Government.
JJB acknowledges the support of an Australian Research Council Future Fellowship (FT180100231).
JvdS acknowledges the support of an Australian Research Council Discovery Early Career Research Award (project number DE200100461) funded by the Australian Government. 
JBH is supported by an ARC Laureate Fellowship FL140100278. The SAMI instrument was funded by Bland-Hawthorn's former Federation Fellowship FF0776384, an ARC LIEF grant LE130100198 (PI Bland-Hawthorn) and funding from the Anglo-Australian Observatory.
The SAMI Galaxy Survey is based on observations made at the Anglo-Australian Telescope. The Sydney-AAO Multi-object Integral field spectrograph (SAMI) was developed jointly by the University of Sydney and the Australian Astronomical Observatory. The SAMI input catalogue is based on data taken from the Sloan Digital Sky Survey, the GAMA Survey and the VST ATLAS Survey. The SAMI Galaxy Survey is funded by the Australian Research Council Centre of Excellence for All-sky Astrophysics (CAASTRO), through project number CE110001020, and other participating institutions. The SAMI Galaxy Survey website is \url{http://sami-survey.org/}.
Parts of this research were supported by the Australian Research Council Centre of Excellence for All Sky Astrophysics in 3 Dimensions (ASTRO 3D), through project number CE170100013.

\section*{Data Availability}
The data used in this manuscript will be made available upon reasonable request to the authors. 

All SAMI data presented in this paper are available from Astronomical Optics’ Data Central service at \url{https://datacentral.org.au/} as part of the SAMI Galaxy Survey Data Release 3.



\bibliographystyle{mnras}
\bibliography{SAMIHIasym} 

\appendix
\section{Extinction corrected global $\HA$ spectra} \label{app:extcorr}
\begin{figure}
    \centering
    \includegraphics[width=0.48\textwidth]{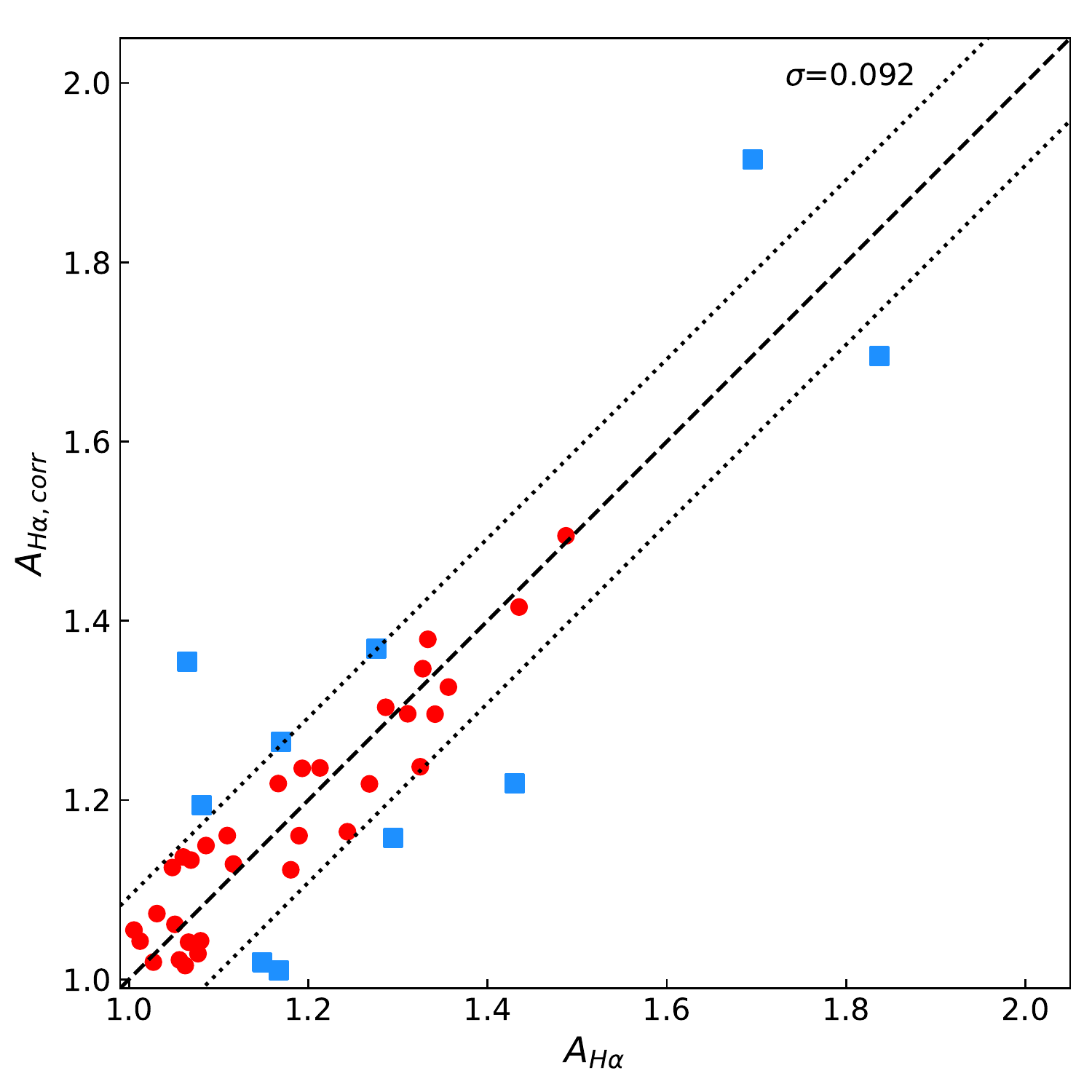}
    \caption{Comparison between uncorrected and extinction-corrected global $\HA$ asymmetries, $\AHA$ and $A_{\HA,\mathrm{corr}}$. The black dashed line is the one-to-one line, and the black dotted lines show the 1$\sigma$ scatter of the $A_{\HA,\mathrm{corr}}-\AHA$ distribution, $\sigma=0.092$. Galaxies outside this 1$\sigma$ scatter are marked with blue squares.}\label{fig:extcorr_HA}
\end{figure}
\begin{figure*}
    \centering
    \includegraphics[width=0.95\textwidth]{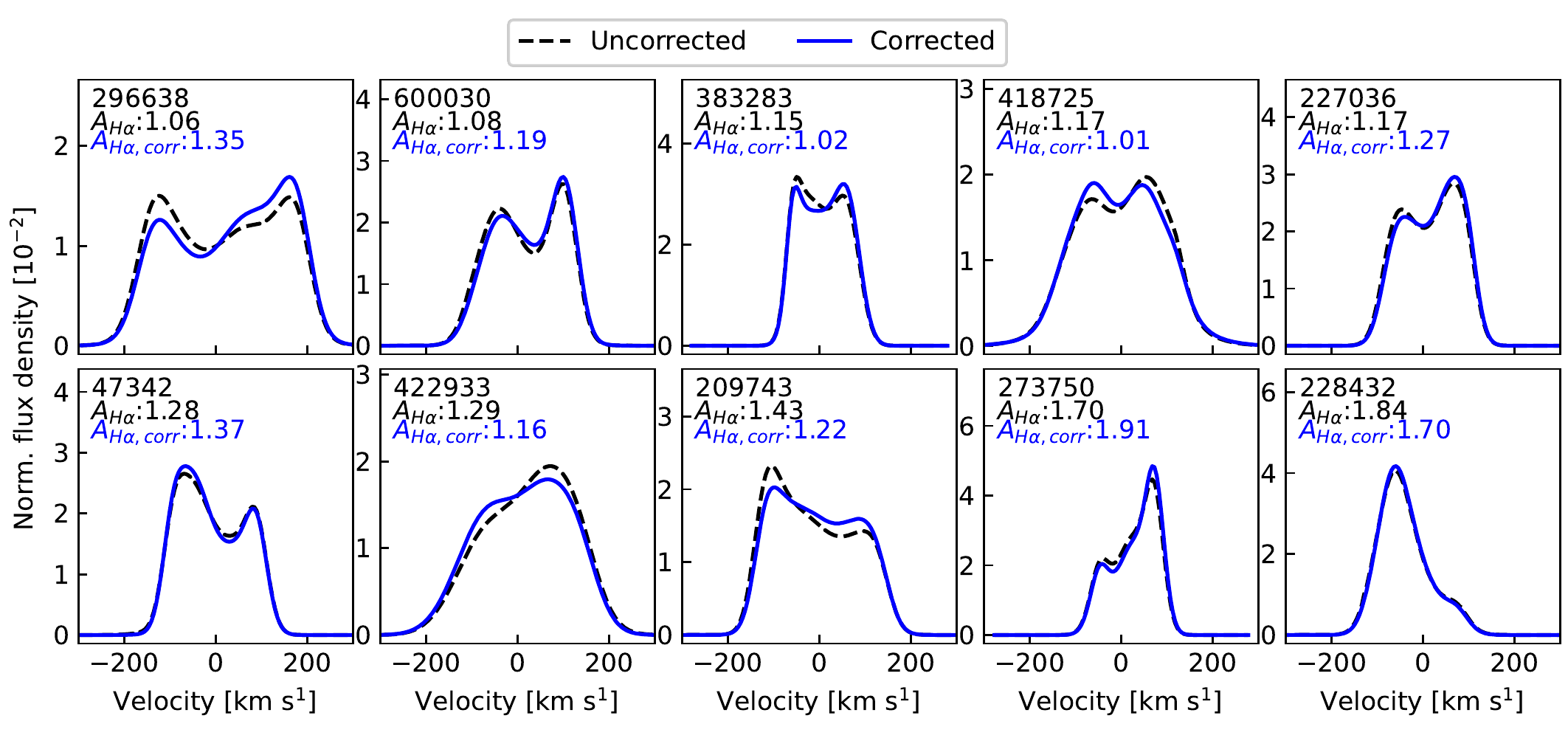}
    \caption{Comparison between uncorrected (black, dashed) and extinction-corrected (blue) global $\HA$ spectra. Both spectra have been normalised to the same integrated flux. The SAMI ID and the values of $\AHA$ and $A_{\HA,\mathrm{corr}}$ are written in the top left corner of each panel.} \label{fig:extcorr_compare}
\end{figure*}
{Here, we investigate the effect of internal extinction on the shape of the integrated $\HA$ profile. 
Using the $\HA$ and H$\beta$ emission line fluxes in the SAMI data product, we derived the extinction E(B--V) assuming a \ct{calzetti00} reddening curve with $R_\mathrm{V}=4$ and an intrinsic  $\HA/H\beta=2.83$ flux ratio. 
We restricted this calculation to all spaxels with $S/N\geq3$ in both lines to ensure that we covered the $S/N_{\HA}\geq5$ requirement for creating our global $\HA$ spectra, and set  E(B--V)$=0$ in spaxels that did not meet this criterion.
We then created global $\HA$ spectra and measured the extinction-corrected asymmetry $A_{\HA,\mathrm{corr}}$ using the same procedure as described in \S\ref{subsec:measure}.}

{In Fig. \ref{fig:extcorr_HA}, we compare $\AHA$ and $A_{\HA,\mathrm{corr}}$ for the sub-sample of SAMI-\HI\ galaxies used in this work.
Clearly, the extinction correction does not significantly change the asymmetry measurement. 
The distribution of $A_{\HA,\mathrm{corr}} - \AHA$ values  has a 1$\sigma$ scatter of $\sigma=0.092$, which we show with black, dotted lines, and galaxies with $|A_{\HA,\mathrm{corr}} - \AHA| \geq 1\sigma$ are highlighted with blue squares.}
{We compare the corrected and uncorrected global $\HA$ spectra of the ten galaxies outside the 1$\sigma$ scatter in Fig. \ref{fig:extcorr_compare}, both normalised to the same integrated flux for better comparison.
Galaxies are ordered by increasing $\AHA$. 
None of the global $\HA$ spectra significantly change their shape, and the main change leading to differences in the asymmetry measurements is a shift in the height of the peaks. 
Further, none of the extinction-corrected spectra would change the classification of whether the $\HA$ and \HI\ spectra match.}

\bsp	
\label{lastpage}
\end{document}